\documentclass[twocolumn,showpacs,pre,superscriptaddress,floatfix]{revtex4}

\usepackage{amsmath}
\usepackage{amssymb}
\usepackage{tabularx}
\usepackage{graphicx}
\usepackage{color}
\usepackage{times}
\usepackage[utf8]{inputenc}

\graphicspath{{./data/}}

\begin{document}

\title{Large-deviation properties of the extended Moran model}

\author{Alexander K. Hartmann}
\affiliation{Institut f\"ur Physik, Universit\"at Oldenburg, D-26111
	     Oldenburg, Germany}
\author{Thierry Huillet}
\affiliation{Laboratoire de Physique Th\'eorique et Modélisation, Universit\'e de Cergy-Pontoise, 
2 Avenue Adophe Chauvin, 95302 Cergy-Pontoise }

\begin{abstract}
The distributions of the \emph{times to the most recent common ancestor}
 $t_{\rm mrca}$ is numerically  studied for an ecological population model,
the \emph{extended Moran model}. This model has a fixed population size $N$.
The number of descendants is 
drawn from a beta distribution Beta$(\alpha, 2-\alpha)$ for various 
choices of $\alpha$. This includes also the classical Moran model 
($\alpha\to 0$)
as well as the uniform distribution ($\alpha=1$).
Using a statistical mechanics-based 
large-deviation approach, the distributions
can be studied over an extended range of the support, down to 
probabilities like $10^{-70}$, which allowed us to study the change of 
the tails of the distribution when varying the value
of $\alpha \in[0,2]$. We find exponential distributions 
$p(t_{\rm mrca})\sim \delta^{t_{\rm mrca}}$ in
all cases,
with systematically varying values for the base $\delta$.
Only for the cases $\alpha=0$ and $\alpha=1$, 
analytical results are known, i.e., 
$\delta=\exp(-2/N^2)$ and $\delta=2/3$, respectively. We recover
these values, confirming the validity of our approach. Finally, 
we also study the
correlations between $t_{\rm mrca}$ and the number of descendants.
\end{abstract}


\maketitle

\section{Introduction}
\label{sec:intro}

Population models \cite{kot2008}
have attracted attention in science for many
decades. One important aspect is to what extent
the genetic variance in a population can be explained without
selection, i.e., by neutral processes. This is also called
\emph{genetic drift}. One very simple
model for this purpose is the well-known Wright-Fisher model 
\cite{wright1931,ewens1979}, where a population of fixed size with
two alleles (variants) of a gene is randomly evolved. In the same
spirit is the Moran model \cite{moran1958}, where no genes are modeled
explicitly, just an inheritance history is considered, which can
be represented by an ancestral tree. 
Note that these models work forward in time. 
Equivalently to the processes forward in time,
 one can construct backward histories, notably within 
so-called \emph{coalescence} models \cite{rosenberg2002}, e.g.,
the \emph{Kingman coalescent} \cite{kingman1982}. An important
feature of the Wright-Fisher and these equivalent models
 is that the individuals have
a small number of offspring, e.g., only zero, one or two in the
extreme case of the Moran model. This leads to long evolutionary time
scales, which are of the size $N$ of the population, if time is measured
relative to $N$, i.e., if the reproduction rate
is constant. 

Nevertheless, there are ecological systems, where
strong fluctuations are observed, in particular in marine 
environments \cite{hedgecock2011} ,
e.g., for populations of oysters \cite{hedgecock1994,eldon2006,sargsyan2008}
or  sardine \cite{matuszewski2018}, but also 
for virus population genetics \cite{irwin2016}. In this case, the number
of offspring has to be modeled by distributions with a large variance
\cite{tellier2014}, often
called \emph{skewed} offspring distributions. 
Thus, one parent might
dominate the next generation, therefore we address this parent here
as \emph{super parent}. When considering Moran-like models, one
often calls this \emph{extended Moran} models. When taking the
coalescence viewpoint \cite{eldon2006}, the corresponding models where
a fraction of all ancestral lines merge are called 
$\Lambda$-coalescent \cite{pitman1999,sagitov1999}, $\Lambda$
denoting the offspring distribution. 
Due to these imposed strong fluctuations, the time scales are much shorter
than for the case of narrow offspring distributions, i.e.,
sub-linear in the population size \cite{eldon2006}. Note that
also from the physics viewpoint it is not surprising that large
fluctuations lead to new effects. This is well known, e.g.,
in the case of phase transitions \cite{stanley1971} where the
fluctuations grow when approaching the phase-transition point.
Nevertheless, in contrast
to the Wright-Fisher and the Moran model, most models with skewed
offspring distributions cannot be solved analytically. Only
for an extended Moran model where the 
offspring distribution is simply uniform in the size
of the population, recently some rigorous 
results were obtained \cite{huillet2013}. Thus, one has to use
numerical simulations of these models if one wants to study them
in the general case.

As for any probabilistic model, one is interested in the behavior
of random variables defined through the model. In the present
work the quantity of interest is the \emph{time
to the most recent common ancestor} (MRCA), 
which describes how fast a population
evolves. For example, the MRCA of humans is estimated to 
have lived some 200,000 years ago, while the MRCA of all present life
lived more than 2 billion years ago \cite{glansdorff2008}.
Often, in exact calculations as well as in numerical
studies, one is restricted to studying averages (or fluctuations) of 
the
quantity of interest. Numerically, a 
rough approximation of the distribution can be
obtained by \emph{simple sampling}: One performs, say, $10^9$ independent
simulations, which allows to obtain the distribution in a range down to
probabilities like $10^{-9}$.
Nevertheless, to obtain a full description of a model, one
would like to obtain the full or at least a much larger 
part of the distribution, which involves often the need to access much
smaller probabilities, i.e., the tails.
This is desirable because, first and in general, 
evolution is often influenced by rare
events. Thus, being able to analyze these rare events allows for
a better understanding of evolution in principle.
Second, more from a fundamental scientific point of view, it is the ultimate
goal to analyze each model as comprehensively as possible. Therefore,
 also within simulations, one desires to obtain any probability distribution
on a range of its support as large as possible, beyond what is possible using
simple sampling. But beyond this scientific interest it is also for
many applications beneficial to obtain at least some part of the tails.
 One important application in the field of the simulation  of
population genetic models \cite{hudson1991}
is hypothesis testing , e.g., to test
whether a neutral model without selection is sufficient to explain
some available data \cite{hudson1994}. Here it is also very useful to
obtain the probabilities down to the range of rather 
small values bey\-ond the limits of simple sampling.

These small probabilities  can actually be reached within
numerical studies by applying
sophisticated so called \emph{large-deviation approaches} 
\cite{denHollander2000,touchette2009,dembo2010,align2002,largest-2011,fBm_MC2013,work_ising2014} which need quite a bit  of additional numerical effort.
They are based on long-time established \emph{importance sampling} or
\emph{variance reduction techniques} \cite{hammersley1956,bucklew2004}.
Often such approaches utilize Markov chain Monte Carlo simulations,
which have been proven to be useful for various other applications in
population genetics \cite{excoffier2006}.
Indeed,
large-deviation techniques have been applied also in population genetics
\cite{stephens2000}, but only for selective questions, not, to our knowledge,
to obtain a probability distribution over a large range of the support.
Here we apply a statistical mechanics-based 
 large-deviation approach to obtain the
distribution of the time
to the most recent common ancestor for a certain class of population models.
This allows us to calculate the distribution in an interval
of the probabilities ranging from the largest of $O(1)$ down
to probabilities as small as $10^{-70}$.

A particular simple class of models
was introduced by Cannings 
\cite{cannings1974, cannings1975}. These  population models
are simple because they exhibit fixed population size 
$\mathbb{N}\in \{1, 2, . . .\}$. These models
are characterized by a family of random variables $\nu_i(t)$, $t \in
\mathbb{Z}=\{\ldots ,-1, 0, 1, \ldots\}$, $i \in  \{1, \ldots , N\}$, 
where $\nu_i(t)$
 denotes
the number of offspring of individual $i$ of generation $t$. Since
we are not interested in the fate of selected individuals, but only
in the evolution of the structure of the population, it is
assumed that for each generation $t$ the offspring variables
$\nu_1(t), \ldots, \nu_N (t)$ are exchangeable, i.e.,
can be permuted.
We consider a particular subclass of the Cannings population
models in which in each generation only one of the
$N$ individuals, the \emph{super parent}, 
is allowed to have more than one offspring. More
precisely, our model is defined in terms of a family of random
variables $U_N (t)$$\in \{0, \ldots, N\}$, 
which denotes the number of offspring of the
super parent in generation $t$. The model is defined
as follows.
For $t \in  \mathbb{Z}$ and $i \in \{1, . . . , N\}$ define
\begin{equation}
\mu_i(t) :=
\left\{
\begin{array}{rl}
1 & \mbox{if } i \in \{1, \ldots , N - U_N (t)\},\\
U_N (t) & \mbox{if } i = N - U_N (t) + 1,\\
0 & \mbox{if } i \in \{N - U_N (t) + 2, \ldots , N\}
\end{array}\right.\,.
\end{equation}
Now let $\nu_1(t), \ldots , \nu_N (t)$ 
be a random permutation of $\mu_1(t), \ldots,\mu_N (t)$. 
For each fixed $t \in \mathbb{Z}$ the random variables 
$\nu_1(t), \ldots , \nu_N (t)$
are then exchangeable and we interpret $\nu_i(t)$ as the number
of offspring of individual $i$ of generation $t$ of a corresponding
exchangeable Cannings model. 

It is assumed that, for each fixed
$N \in \mathbb{N}$, the random variables $U_N (t)$, $t \in \mathbb{Z}$, 
are independent and
identically distributed. 
 The most celebrated example is the standard Moran
model \cite{moran1958}  corresponding
to $U_N \equiv 2$, in which the super parent  has two
offspring, one other randomly selected individual has no offspring
and all the other $N - 2$ individuals have exactly one offspring.

Here, we consider an extended Moran model \cite{huillet2013}
for the case where $U_N=Nr$ and the random number 
$r\in[0,1]$ is drawn from
a beta-distribution Beta($\alpha, 2-\alpha$), i.e.,
with density $r^{\alpha-1}(1-r)^{1-\alpha}/(\Gamma(\alpha)\Gamma(2-\alpha))$.
The ``reproduction parameter'' $\alpha$
allows us to interpolate
between the case where the super parent has only a small number of offspring
($\alpha\to 0$) and between the case where the super parent takes over
the population very quickly ($\alpha\to 2$). Also the case of a uniform
distribution is included ($\alpha=1$). 
Such a process has also been studied \cite{bolthausen1998} in physics in the
context of disordered magnetic systems called spin glasses
\cite{binder1986,mezard1987,young1998}. 
The case of the beta 
distribution for $\Lambda$-coalescents has been introduced
before \cite{schweinsberg2003} and
is sometimes called beta-coalescent \cite{tellier2014}.

\section{Algorithms}

In the first of the following two subsections, 
we explain how we simulated the extended Moran model
in order to measure the time $t_{\rm mrca}$ to the most recent common ancestor.
This is pretty straightforward. 

To obtain the distribution in ranges where the
probabilities are as small as $10^{-70}$,
we used a previously-developed
large-deviation algorithm presented below in the second subsection.
It consists of a separate level of a Markov Chain Monte Carlo (MCMC) 
simulation, on top of the simulation of the Moran model.

Here we present only a general outline of the algorithm, and
the details which are specific to the simulation of the extended
Moran model. Nevertheless, the algorithm works in a general 
way such that it can be applied
to the simulation of an arbitrary ``target'' 
stochastic process. Each instance of a simulation
 of the target process is assumed
to result in a measurable scalar quantity $W$ of interest,
exhibiting a probability distribution $P(W)$ (below $W$
will be the time $t_{\rm mrca}$ to the most recent common ancestor). 
On a digital computer, stochastic processes
can be simulated using (pseudo) random numbers, denoted here as $\{\xi_i\}$.
 Usually, the (pseudo) random number are computed on the fly while the target 
simulation is 
performed. Equivalently, one can \emph{precompute} (or obtain in a
different way) the random numbers before the actual simulation is 
performed \cite{crooks2001}.
This  \emph{set} of random numbers is stored in 
a vector $\xi=(\xi_0,\xi_1,\ldots,\xi_{M-1})$
of suitable length $M$. While the actual target process
is simulated, the random numbers used are taken from the set $\xi$.
Therefore, ignoring the dependence on initial conditions etc, the
outcome $W$ of the target process 
depends only \emph{deterministically} on the set of used randoms numbers,
i.e., $W=W(\xi)$. Most general, the entries of 
$\xi$ are random variables uniformly distributed in $[0,1]$, 
since any type of random numbers can be obtained from them via 
suitable transformations. 

Note that, to ensure a good convergence behavior of the algorithm,
 one should use each entry of $\xi$
always for the same purpose, independently of other entries of $\xi$.
This means, as we will see below, 
some values of $\xi$ will be ignored sometimes.
This helps to ensure that a small change in $\xi$ leads typically
to a small change of $W$, which is necessary for a good behavior of
the algorithms used below.

\subsection{Simulating the extended Moran model\label{sec:moran_algorithm}}

As introduced above, we assume that the random numbers, 
uniformly $U(0,1)$ distributed,
  needed for performing one run (population
size $N$ and $t_{\max}$ generations) are stored in a vector 
$\xi=(\xi_0,\xi_1,\ldots,\xi_{M-1})\in [0,1)^M$ of size $M=t_{\max}\times(N+2)$.
Thus, to simulate a single generation (up to) $N+2$ random numbers are
needed. 

Here, the evolution starts with the population at time $t=0$.
 The evolving fate of the population is stored as vectors 
$a(t)~=(a_1(t),\ldots, a_N(t))$ of numbers denoting
the corresponding  ancestors at the initial time $t=0$. 
Thus, the value $a_i(t)$ denotes  
in generation $t$ the ancestor of individual $i$ from generation 0. Therefore,
the vector is initialized as $a_i(0)=i$ for $i=\{1,\ldots,N\}$.

The evolution of the population is simulated forward in time.
In each generation $t=1,\ldots,t_{\max}$, 
one super parent $n_0\in\{1,\ldots,N\}$  is selected randomly and uniformly
in the population. For this purpose, the entry $\xi_{t(N+2)}$ is
used. Thus, the random selection is achieved via simply setting 
$n_0=\lfloor N \xi_{t(N+2)} \rfloor$.
Next, the number $U_N(t)$ of offspring of $n_0$ is selected via
drawing a random number from the Beta($\alpha, 2-\alpha$) distribution.
Drawing this random number works using
the inversion method \cite{gsl2006}, for this purpose the entry
  $\xi_{t(N+2)+1}$ is used.
The generated number is multiplied by $N+1$ and the floor is taken, 
resulting in $U_N(t)\in\{0,\ldots,N\}$. Note that if $U_N(t)=0$ we
define that still
individual $n_0$ will generate one offspring by the next step:
From the population at time $t-1$ those $N-U_N(t)$
members $n$ have to be selected (uniformly), which have exactly one offspring,
i.e., $a_{n}(t)=a_{n}(t-1)$.
For this purpose the entries
$\xi_{t(N+2)+2},\ldots,\xi_{t(N+2)+1+N-U_N(t)}$ are used,
i.e., none if $U_N(t)=N$. Thus, the 
subsequent entries
$\xi_{t(N+2)+1+N-U_N(t)},\ldots,\xi_{(t+1)(N+2)-1}$
are not used for the simulation. Finally, the members $n$ of the 
population, which were
\emph{not} among those selected for having exactly one offspring, 
are determined
as  offspring of individual $n_0$, i.e., $a_n(t)=a_{n_0}(t-1)$.
By this, the calculation of generation $t$ is complete.

\begin{figure}[ht]
\begin{center}
\includegraphics[width=0.7\linewidth]{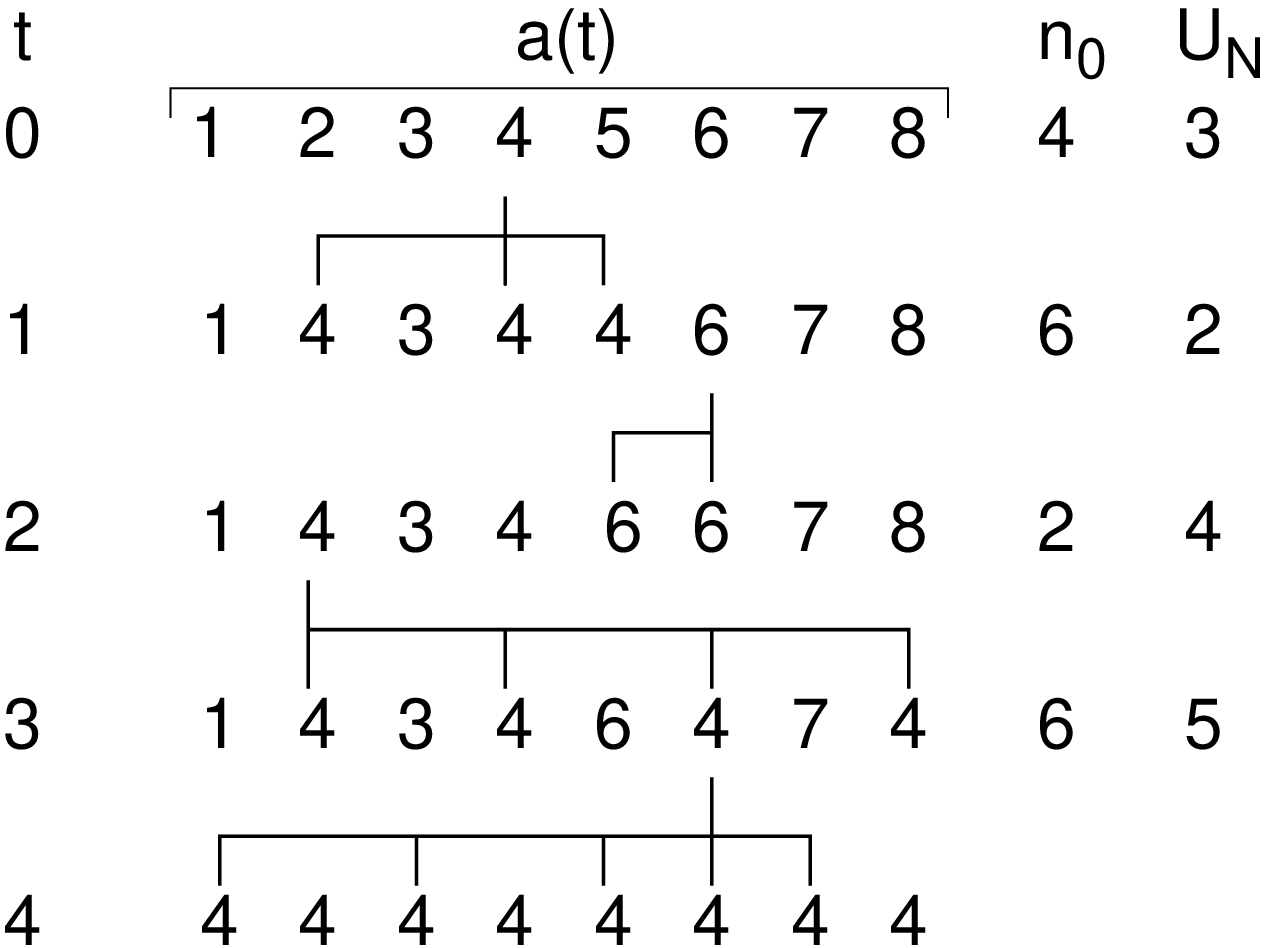}
\end{center}
\caption{Sample evolution for a population with $N=8$ individuals. 
Shown are the number $t$
of generations (left), the values of the ancestors $a_1(t),\ldots,a_N(t)$,
the selected super parent $n_0$ for the next generation, and the
number $U_N(t)$ of offspring of the super parent. In this case,
 after four generations, all individuals of the populations are
descendants of individual 4 of the initial generation.
\label{fig:evolution}}
\end{figure}

In Fig.\ \ref{fig:evolution} one sample evolution is shown. As visualized
in the figure, at some random time, 
for a finite population size $N$
and if the total simulation time $t_{\max}$ is large enough,
all members population are for the first time
offspring of the same individual which was present in the initial population
($t=0$). Thus, they all have a common ancestor, i.e., 
$a_n(t_{\min})=a_1(t_{\min})$ for $n=1,\ldots,N$. We say the individual
\emph{dominates} the population.
Thus, looking from this time backward to the initial 
configuration, which is statistically equivalent to looking forward,
this time is the time $t_{\rm mrca}$ to the most recent
common ancestor we are interested in, i.e., $t_{\rm mrca}=t_{\min}$

Note that we measure the probability $P(t)$ that {\em any}
individual dominates (for the first time) after $t=t_{\min}$ steps, which
means we look forward in time. 
This probability is interestingly the
same as the probability 
$P(i\, {\rm dom. \,after\,} t|i\, {\rm dom.})$ 
that a \emph{specific} (say $i=1$) individual dominates 
after $t$ steps, conditioned that it is individual $i$ which dominates
after some time, which basically means one starts at the time where
one specific individual dominates, and looks backwards in time. This
can be seen easily, because:

\begin{eqnarray}
P(t) &=& 
\sum_{i} P(i\, {\rm dom. \,after\,} t|i\, {\rm dom.}) 
P(i\, {\rm dom.}) \\
&=& N P(1\, {\rm dom. \,after\,} t|1\, {\rm dom.}) \frac 1 N \\
&=&  P(1\, {\rm dom. \,after\,} t|1\, {\rm dom.})\,
\end{eqnarray}
because by symmetry of all individuals $P(i\, {\rm dom.})=\frac 1 N$
and the conditional probabilities are all the same, thus the sum reduces
to a multiplication with $N$.

One can measure, for example, the mean of $t_{\rm mrca}$ and obtain a 
(small-support) histogram
via \emph{simple sampling}: One generates, say, $K$ times a vector $\xi$
of random numbers, and runs each time the above described algorithm to generate
the dynamics
 of the evolution. Then one measures for each run the resulting
time where for the first time all members of the population originate from
the same ancestor. Depending on the value $K$ of independent runs,
the obtained histogram will be of better or worse quality. Typically,
Probabilities of the order of $1/K$ can be measured, like $10^{-9}$.

\subsection{Large-deviation approach\label{sec:large_deviation}}

Following the description so far, one is able to simulate the evolution
of the population in a standard way. The only difference
to standard implementations is that the
generation of the random numbers and the actual simulation of the
stochastic target process are disentangled.

Nevertheless, this disentanglement allows one to perform a Markov-chain
Monte Carlo evolution for the set $\xi$: $\xi^{(0)}\to\xi^{(1)}\to
\xi^{(2)}\to\ldots$. Thus, the target process is not performed several
times independently, but for a sequence of similar (correlated) sets
$\xi^{(s)}$ ($s=0,1,2,\ldots$).
 This may appear inefficient on the first sight since the
measured quantities $W(\xi^{(0)})$, $W(\xi^{(1)})$, $W(\xi^{(2)})$, \ldots 
will be correlated as well, in contrast to using each time a new
set $\xi$, corresponding to \emph{simple sampling}.
 On the other hand, the MC evolution allows one to introduce
a bias, such that the measured distribution for $W$ can be directed
into a region of interest, e.g., where the original probabilities $P(W)$ are
very small.  Here, since we are inte\-res\-ted in the distribution of $W$
for a large range of the support, an exponential Boltzmann bias
$e^{-W/\Theta}$ is used where  $\Theta$
is a freely adjustable parameter (a kind of artificial temperature), 
which allows us to center the observed distribution in
different regions.   Note that the choice $\Theta=\infty$ corresponds
to the absence of the bias, i.e., to the simple sampling presented
in Sec.\ \ref{sec:moran_algorithm}, only including correlations.
The bias is included in a standard Metropolis-Hastings
algorithm where vectors $\xi^{(s)}$ ($s=0,1,2,\ldots$)
are sampled via changing a small fraction of entries of the
current vector in each step.
For details, please consider Ref.~\cite{work_ising2014}.

Note that this approach does not need any
detailed knowledge about the specific model simulated, which
is often used to ``guide'' a large-deviation approach to the
region of interest. Thus, the approach is
in contrast 
to many standard large-deviation algorithms for 
dynamical equilibrium and non-equilibrium systems 
\cite{berryman2010}. Within the present approach the
target process leading to the measurable quantity $W$ can be seen
as a black box. This allows one 
to perform large-deviation measurements
for almost arbitrary equilibrium and non-equilibrium processes
which can be simulated on a computer using (pseudo) random numbers.

\begin{figure}[ht]
\begin{center}
\includegraphics[width=0.9\linewidth]{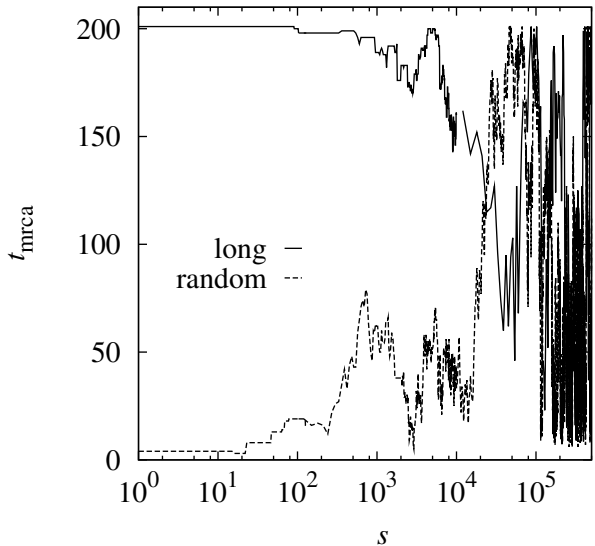}
\end{center}
\caption{Time $t_{\rm mrca}$ to most recent common ancestor as 
function of the Monte Carlo time $s$ for population size $N=200$,
reproduction parameter $\alpha=1$, 
sampling temperature $\Theta=-2.5$  for two different initial
start configuration $\xi^{(0)}$: one completely
random and one with a bias towards larger values of $t_{\rm mrca}$.
\label{fig:equilibration}}
\end{figure}

In Fig.\ \ref{fig:equilibration} two sample evolutions of $t_{\rm mrca}$
are shown as a function of the number $s$ of Monte Carlo steps. Both
evolutions are for a population size $N=200$, reproduction parameter
$\alpha=1$ and MC temperature $\Theta=-2.5$. The negative temperature
results in values of $t_{\rm mrca}$ which are larger than typical values
for this choice of $N$ and $\alpha$. Note that the two data sets
start from opposite sides: one set is for a initial configuration vector
$\xi^{(0)}$
which is drawn independently from $[0,1)^M$. This corresponds to the typical
behavior. The other vector is drawn such that it exhibits a preference
for large values of $t_{\rm mrca}$. This is achieved by sampling the entries
which are responsible for drawing the number of offspring not uniformly
in the interval $[0,1]$ but in a smaller interval $[0,0.15]$. In spite
of the different initial conditions, after a while the value of the
measured quantity $t_{\rm mrca}$ agree within the range of the fluctuations.
This indicates that the Markov chain has ``forgotten'' its initial condition,
i.e., can be considered as \emph{equilibrated}.

For each value of $\Theta$, one obtains a distribution which
includes the original distribution under the bias $e^{-W/\Theta}$.
Such biased simulations have been performed already in the field
of population biology, e.g., to estimate small likelihoods \cite{stephens2000}. 
Nevertheless,
here we are interested in obtaining the distribution of interest for a
large range of the support, which has not been done in population genetics
to our knowledge. To achieve this we have to perform the simulations
for \emph{several} values of $\theta$ and combine the results
in a suitable way, as shortly outlined now. The sampled distribution at 
value $\theta$ 
is related to the original distribution \cite{align2002} via
\begin{equation}
P(W)=e^{W/\Theta}Z(\Theta)P_{\Theta}(W)
\label{eq:PW}
\end{equation} 
where $Z(\Theta)$ is the normalization constant of $P_{\Theta}(W)$,
which can be determined from the numerical data, as explained next.
By performing the simulation for suitably chosen values of $\Theta$,
which have to be determined experimentally, 
one can cover a broad range of the desired distribution $P(W)$.
If this is done such that
the resulting distributions $P_{\Theta}(W)$ overlap for neighboring
values of the temperature, say $\Theta_1$ and $\Theta_2$, one can reconstruct
the relative normalization constants from 
$e^{W/\Theta_1}Z(\Theta_1)P_{\Theta_1}(W)= e^{W/\Theta_2}Z(\Theta_2)P_{\Theta_2}(W)$.
Actually, several values of $W$ in the windows 
$[W^{\min}_{\Theta_2}, W^{\max}_{\Theta_1}]$
of overlap are considered
and the mean-square difference in this window 
between the distributions $P(W)$ obtained
from (\ref{eq:PW}) is minimized to obtain the ``optimum'' relative
normalization constant $Z(\Theta_1)/Z(\Theta_2)$. 
The last missing
constraint is obtained from the overall normalization of $P(W)$ which
then results in the actual values for all normalization constants.
 Details can be found in Ref.\ \cite{align2002}, a generalization is explained
in Ref.\ \cite{shirts2008}.

\section{Results \label{sec:results}}

We have performed simulations for different population sizes $N$,
different values of the reproduction parameter $\alpha$ and various
temperatures $\Theta$. The number of different temperatures for a given
combination of $N$ and $\alpha$ ranged from two to seven.

For each combination of these parameters,
the number of how many of the entries 
of the configuration vector $\xi$ were redrawn in each MC step
to obtain the trial
vector was adjusted. As a rule of thumb
(often used for Monte Carlo simulation), this amount of adjustment
 was chosen such that the acceptance rate of the
Metropolis steps was roughly near 0.5.
Note that the number $c_{\rm sp}$ of changes for the entries of $\xi$ 
which are responsible for selecting the super parent,
for selecting the number of offspring ($c_{\rm o}$), 
and for selecting the one-offspring  parents ($c_{\rm 1}$), 
were adjusted separately. 
The reason is that the amount of change for 
super parent
entries and numbers of offspring entries 
have a higher influence on the acceptance rate than
the amount of change for the one-offspring parent entries. 

\begin{figure}[ht]
\begin{center}
\includegraphics[width=0.9\linewidth]{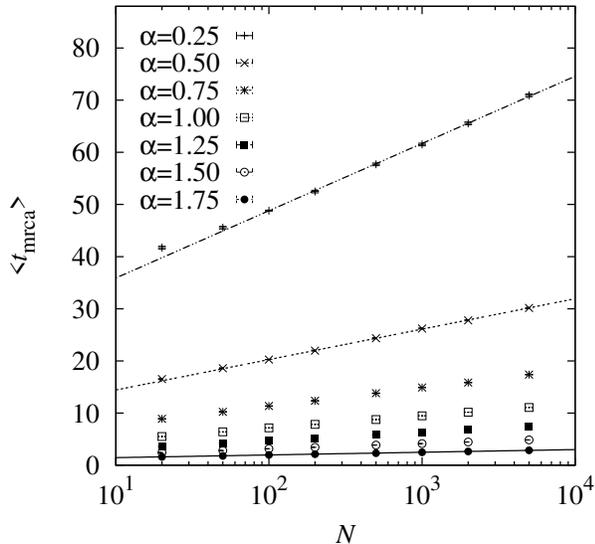}
\end{center}
\caption{Mean time $t_{\rm mrca}$ to most recent common ancestor as 
function of the population size $N$ for different values of the 
reproduction parameter $\alpha$.
\label{fig:mean_tmrca}}
\end{figure}

Using simple-sampling, we measured the mean of the time 
$t_{\rm mrca}$ to the most recent common
ancestor as a function of the population size $N$.
 The result is shown in Fig.\ \ref{fig:mean_tmrca} using a 
logarithmically-scaled $N$-axis. When ignoring very small values of $N$,
the data follows straight lines very well, meaning that
it is described well by a logarithmic growth.
This matches an analytical calculation for the 
uniform distribution, i.e., the case $\alpha=1$ \footnote{The
mean is stated in general in Ref. \cite{huillet2012} after Eq. (37), see
page 121. For the
present case one has to perform a short calculation by using
 the corresponding measure which results
in a $\log(N)$ dependence.
very well. Also for $\Lambda$-coalescent, a fast evolution is obtained
\cite{eldon2006}.}.

\begin{figure}[ht]
\begin{center}
\includegraphics[width=0.9\linewidth]{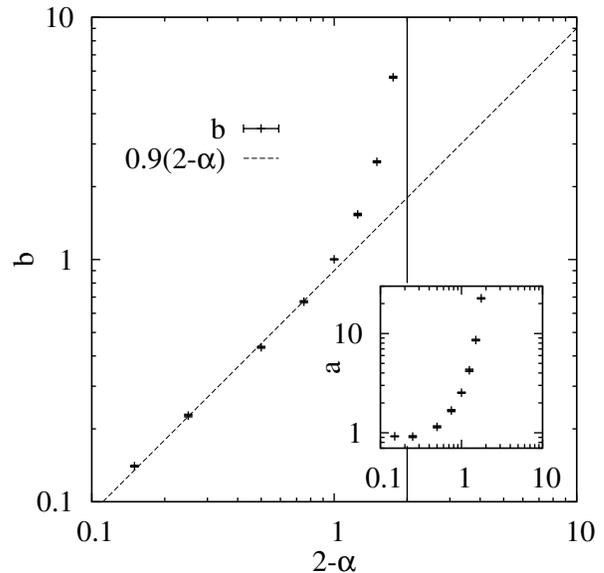}
\end{center}
\caption{Fit parameter $b$ when fitting $t_{\rm mrca}(N)=a+b\log(N)$
 as function of the population size $N$. The resulting value of $b$
is shown as a function of  $2-\alpha$
(with $\alpha$ being the reproduction parameter).
The vertical line
indicates the end of the range of feasible values for $\alpha$. The dashed line
approximates the behavior for values of $\alpha$ close to $2$ and serves
as guide to the eyes.
 The inset shows the behavior of fit parameter $a$
as function of $2-\alpha$.
\label{fig:const_mean_fit}}
\end{figure}

With increasing value of $\alpha$ the typical number of descendants
increases. Thus, one can expect that the mean $\langle t_{\rm mrca}\rangle$ 
decreases
when $\alpha$ increases. For the same reason one can expect that for 
$\langle t_{\rm mrca}\rangle \sim a+b\log N$ the slope value $b$ decreases  
when $\alpha$ increases. We have fitted the data to this functional
form.
The result is shown in Fig.\
\ref{fig:const_mean_fit} as a function of $2-\alpha$. 
Note that
for $\alpha=1$ we obtained $b=1.002(4)$ which is compatible with $b=1$.
In the log-log
plot, a straight line is visible for $\alpha\to 2$, meaning that $b$
converges to zero. Comparably, $a$ appears to converge to a finite value 
close to 1 (see inset of Fig.\ \ref{fig:const_mean_fit}).
This is reasonable because for $\alpha\to 2$ the
distribution of the number of descendants converges to the case
where the super parent takes over the full population in one generation,
independent of the size of the population,
which means $t_{\rm mrca}=1$ with $a=1$ and $b=0$.

In the limit $\alpha\to 0$, $b$ diverges, i.e., the time grows more
than logarithmically. This appears also to be reasonable, because
in this limit the model converges to the Moran model, being equivalent
to the Kingman coalescent, where $t_{\rm mrca} \sim N^2$ for the time
scale we use ($\sim N$ if the time is rescaled by $N$ as in the standard
studies). Thus, in the general case $\alpha>0$, the evolutionary 
timescale grows much slower with increasing population size. This 
allows, as in the case of marine environments, 
for a much faster adaption to changing environments, if, e.g.,
selective pressure is present.

\begin{figure}[ht]
\begin{center}
\includegraphics[width=0.9\linewidth]{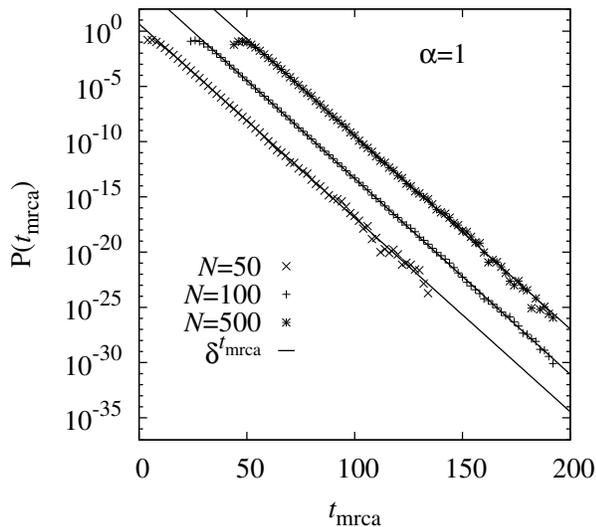}
\end{center}
\caption{Distribution of the time $t_{\rm mrca}$ to the most recent common 
ancestor for reproduction parameter $\alpha=1$ and three
different population sizes $N=50$, $N=100$ and $N=500$. For better
visibility, the data for $N=100$ is shifted by 20 time units to the
right, and the data for $N=500$ by 40 time units. The lines
show the results of fitting the tails of the distributions
 to exponential functions, respectively.
\label{fig:distr_a1_N}}
\end{figure}

Next, we study the full distribution of $t_{\rm mrca}$,
see Fig.\ \ref{fig:distr_a1_N} for the case $\alpha=1$, 
and different population sizes $N$. This value corresponds
to the uniform distribution and only for this case of a skewed
offspring distribution for the extended Moran model rigorous results 
\cite{huillet2013}
are available to which we can compare.
Using the large-deviation approach, we could measure the
 distribution over a large range of the support.
 Apparently these tails exhibit an exponential shape. 
We  fitted these tails to functions
$\sim \delta^{t_{\rm mrca}}$ for all values of $N$.
We obtained $\delta=0.668(2)$ ($N=50$), $\delta=0.665(1)$ ($N=100$),
$\delta=0.668(1)$ ($N=500$) (The error bars are just statistical
error bars).
 All these values are very close to the exact value \footnote{See bottom
of page 145 of Ref.\cite{huillet2012}.}
$\delta(1)=2/3$  and do within
the fluctuations not depend  on the population size. Thus,
even small population sizes are suitable for obtaining results
close to the $N\to\infty$ limit, if the tails
of the distribution are accessible. For this reason, we proceed with results
for $N=100$, for various values of $\alpha\in[0,2]$.

\begin{figure}[t]
\begin{center}
\includegraphics[width=0.9\linewidth]{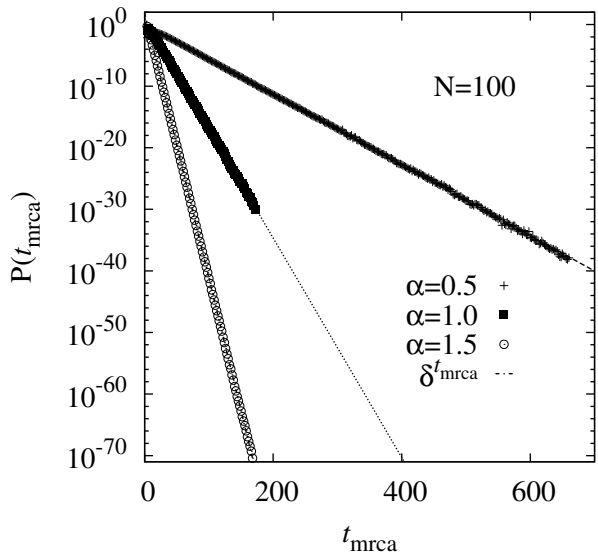}
\end{center}
\caption{Distribution of the time $t_{\rm mrca}$ to the most recent common 
ancestor for population size $N=100$ and three example values of the 
reproduction parameter $\alpha$.
\label{fig:distr_f1_N100}}
\end{figure}

We have performed extensive simulations for other
values of $\alpha$, where no rigorous results are available. Since
we were able to reproduce rigorous results for the test case $\alpha=1$,
we are very confident that our results are reliable also
for these other values of $\alpha$.
In Fig.\ \ref{fig:distr_f1_N100} the distributions for three different
values of $\alpha$ and a population size of $N=100$ are shown.
Again, the tails of the distributions can be well fitted to exponential functions
$\sim \delta^{t_{\rm mrca}}$ for all values of $\alpha$. We did this
for all values of $\alpha$ which we have studied. 


\begin{figure}[!htb]
\begin{center}
\includegraphics[width=0.9\linewidth]{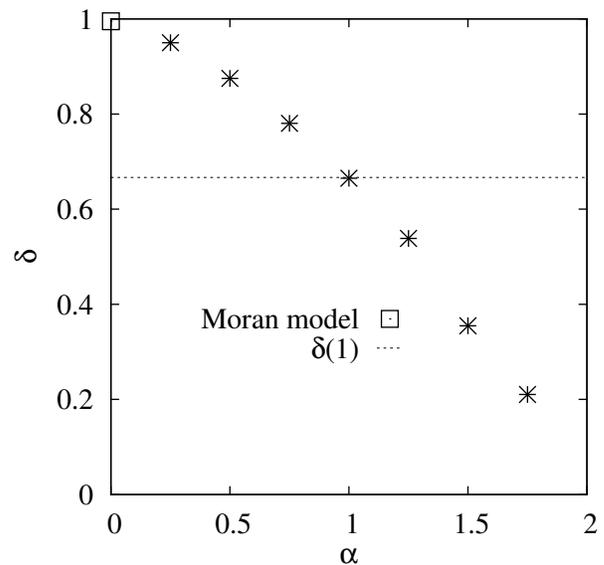}
\end{center}
\caption{Value of the base $\delta$ of the tail 
$P(t_{\rm mrca})\sim \delta^{t_{\rm mrca}}$
of the distribution of the time to the most recent common ancestor 
as a function of the  reproduction parameter $\alpha$. For $N=100$.
The two values which are exactly known are indicated in the plot:
For the standard Moran model (corresponding to $\alpha\to 0$), the full
distribution is know analytically, here 
$\delta=\exp(-2/N^2)$, which is shown as square symbol. 
For the case $\alpha=1$ not the full distribution but the
tail behavior is known analytically, Here a  horizontal line
indicates the exact value $\delta=2/3$.
\label{fig:base_tmrca}}
\end{figure}

In Fig.\ \ref{fig:base_tmrca}, the resulting behavior of $\delta$
as a function of $\alpha$ is shown. With increasing value of $\alpha$, i.e.,
when the distribution of the number of offspring of the super parent is more
and more located at large values, the value of $\delta$ decreases, 
corresponding to smaller times it takes for one individual to dominate the
full distribution. 
Thus, for $\alpha\to 2$ we obtain $\delta\to 0$ corresponding to an 
$t_{\rm mrca}=1$.

For the opposite limit $\alpha\to 0$ the model converges to the 
classical Moran
model \cite{moran1958}, this corresponds to the Kingman
coalescent for the $\Lambda$-coalescent. Here, the exactly-known 
distribution of $2t_{\rm mrca}/N^2$
follows (see Eq. (53) and below in Ref.\ \cite{huillet2016}) 
in the tails a simple exponential distribution  $\exp(-t)$.
This corresponds to a value $\delta=\exp(-2/N^2)$ which evaluates for
the present case ($N=100$) to $\delta=0.995$, i.e. very close to 1,
as shown in the figure. This means the distribution becomes,
without rescaling the time,
infinitely long stretched for the Moran model in the case $N\to\infty$.
 On the other hand,
the extended Moran model for apparently all values of $\alpha>0$,
even the smallest ones,  shows a completely different
behavior, all distributions fall off exponentially even in the limit $N\to
\infty$. This corresponds to the slow logarithmic growth of the average
$t_{\rm mrca}$.

\begin{figure}[ht]
\begin{center}
\includegraphics[width=0.9\linewidth]{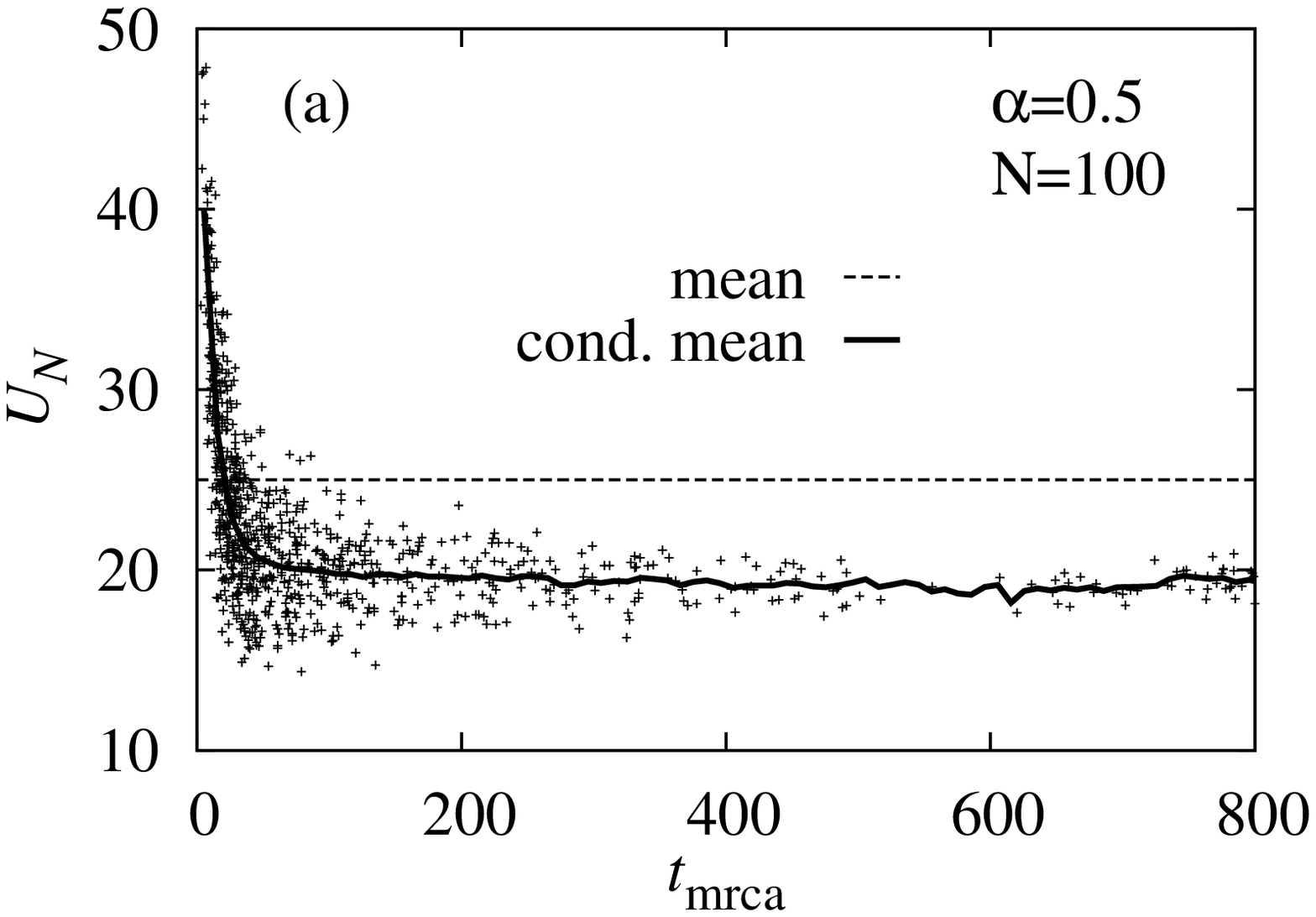}
\includegraphics[width=0.9\linewidth]{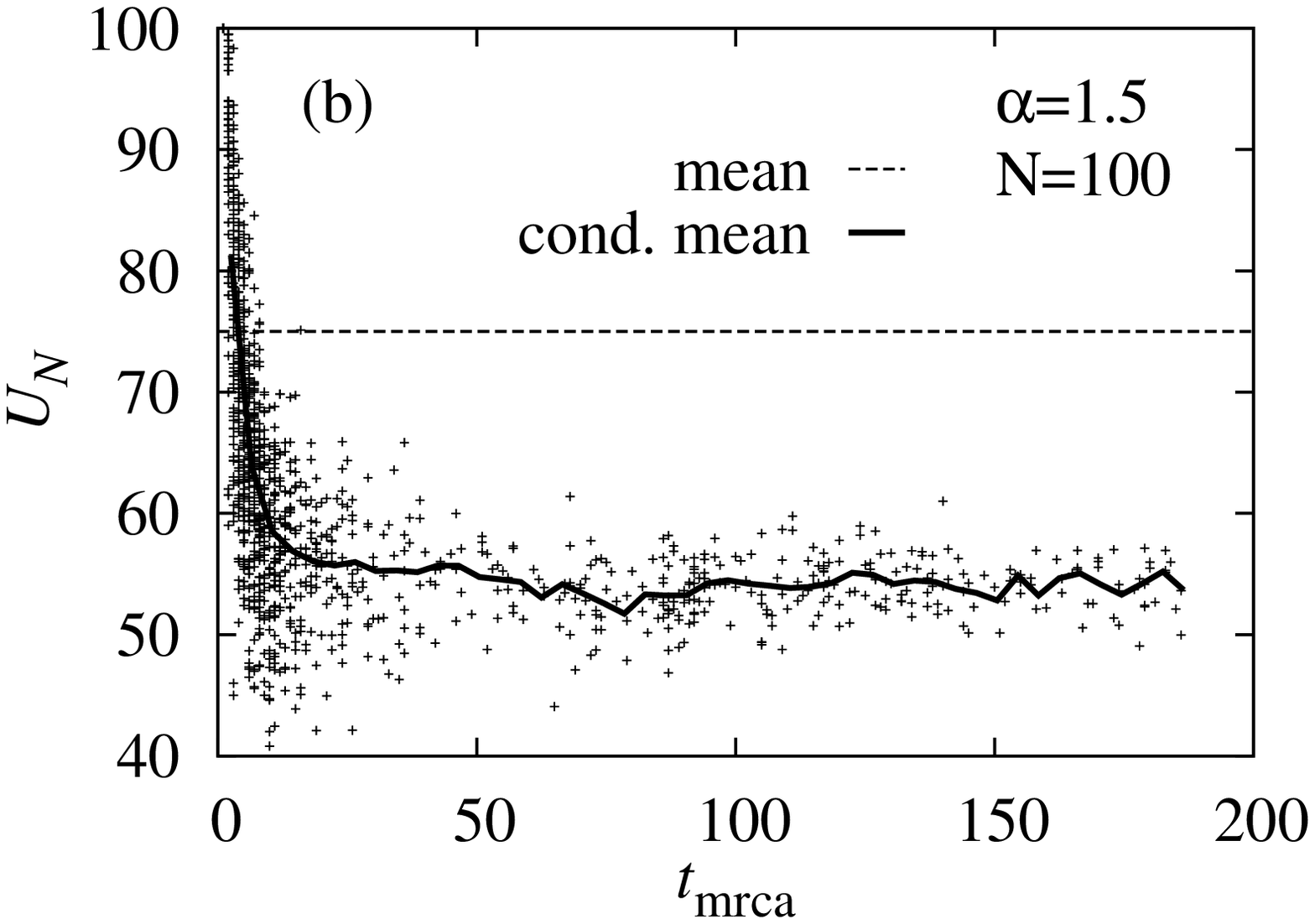}
\end{center}
\caption{Correlations between the $t_{\rm mrca}$ and the average
number $U_N$ of offspring generated during an evolution for
(a) $\alpha=0.5$  and (b) $\alpha=1.5$. The small symbols
present a scatter plot of samples. The horizontal broken line shows 
the unconditioned mean of $U_N$ as given by the Beta distribution,
i.e., $\alpha N/2$. The solid line shows the average of $U_N$ 
conditioned to those evolutions which exhibit a particular
value of $t_{\rm mrca}$.
\label{fig:correlation}}
\end{figure}

Finally, we want to understand what leads during the
evolution of a population to particular small or 
large value of $t_{\rm mrca}$. For this purpose, we study the correlation
between the average value of the number $U_N$ of descendants
encountered during an evolution and relate it to the value of $t_{\rm mrca}$.
As shown in Fig.\ \ref{fig:correlation}, very small values of $t_{\rm mrca}$
correspond to untypical large values of the number of descendants, while
slow evolutions, i.e., large values of $t_{\rm mrca}$
correspond typically to small number of descendants. Note that the full
range of correlations is only accessible to us because we used the
large-deviation approach. The high-probability simple-sampling range
is only within the first 10\% of the range of values for $t_{\rm mrca}$.
The observed behavior is somehow
expected because many decedents lead clearly to a faster evolution of  the
population. The general result
holds for all values of $\alpha$. One is tempted to assume that 
extreme
long evolutions to the MRCA are created by periods with even
smaller number of  descendants.
This could be caused in real environments, e.g., 
by long periods of low but not extremely low availability of nutrition.
Interestingly,
in the tail of the distribution, i.e., for very large 
values of $t_{\rm mrca}$, the values of $U_N$ drop only
very slightly when increasing $t_{\rm mrca}$. This means, extreme
long evolutions to the MRCA are rather an effect of rare choices
of the particular sets of descendants, with a still considerable large number
of descendants. 
Also, the shapes of the scatter plots look very similar for all
values of the reproduction parameter $\alpha$. For the smaller value of 
the reproduction parameter, the variation of $U_N$ is smaller and 
farther away from the unconditional expectation value $\alpha N/2$.
This also appears to be reasonable, because the distribution $P(t_{\rm mrca})$
falls off much quicker for small values of $\alpha$, thus $U_N$
cannot vary much during an evolution such as to result in a long evolution
until the MRCA.

\section{Summary and Discussion}

We have studied the extended Moran model, where possibly
a large fraction of the population is exchanged between two
generations. This model is suitable, e.g.,
to describe in a simplified way 
the population dynamics in marine environments.
We have studied a 
Beta($\alpha,2-\alpha$)-distribution for the fraction of the population
which descends from the super parent, as introduced
in Ref.\ \cite{schweinsberg2003}. The variation
of the parameter $\alpha$ covers a large range of possible distributions,
including the Moran case, the uniform distribution and the limit of a
complete exchange of the full population. 
In particular we have studied
the time $t_{\rm mrca}$ to the most-recent common ancestor,
which describes the timescale on which the genetic drift
takes place. The
typical behavior (like the mean) of this quantity can be readily studied.

Nevertheless, in order to describe the statistics of this model 
as comprehensively as possible,
we investigated not only the typical behavior but also the distribution
of $t_{\rm mrca}$ for a range of the support as
large as possible, including very small probabilities.
 Here, analytical results
are only available for the cases of a uniform distribution, i.e.,
$\alpha=1$, and for the case $\alpha\to 0$ which is the original
Moran model. To access these distributions numerically for arbitrary
values of $\alpha$, we used an established but elaborate
 statistical mechanics-based biased sampling approach. It  
is based on a Markov chain evolution of a vector of
uniformly distributed random numbers from the interval $[0,1]$, 
seen as an input vector to an arbitrary stochastic process.

We found that the mean time $t_{\rm mrca}$ depends for all values 
$\alpha>0$
logarithmically on the population size and converges to a constant for
$\alpha\to 2$. This is in contrast to the 
quadratic behavior of the standard Moran model, 
and reflects a much faster evolution of
a population, as observed for
studies of the $\Lambda$-coalescent and in previous
experimental work on marine populations.

The distribution of $t_{\rm mrca}$ shows an exponential
behavior in the tails. For the cases $\alpha=1$ and $\alpha=0$ 
(where the ``exponential'' becomes infinitely strongly stretched
for $N\to\infty$)
both mean as well as tail behavior are compatible with previous analytical 
results. By studying the correlations of $t_{\rm mrca}$ to the
history of the evolutions, we could show that medium rare deviations
are caused by unlikely small numbers of descendants, while the strongest
deviations are caused by a combination of small but not too small
numbers of descendants
and unlikely combinations of selected offspring.

This work also shows, in line with previous applications
in other fields,
 that using sophisticated sampling techniques, 
the distributions of measurable quantities in 
population models can be studied over large ranges of the support.
This allows one to access results in regions where no analytical
results are available, as here for the extended Moran model. It would be
interesting to apply such techniques to more refined models,
such models with varying population size, or including selection,
or models having a spatio-temporal evolution.

\begin{acknowledgments} 
We thank Charlotte Lotze and Hendrik Schawe for critically reading the manuscript.
AKH is grateful to the Universit\'e Cergy-Pontoise for financing 
a two-month visit in 2016, during which the main parts of the project
were performed. 
The simulations were performed on the HPC clusters HERO
and CARL  of the University
of Oldenburg jointly funded by the DFG 
through its Major Research Instrumentation Programme (INST 184/108-1 FUGG
and INST 184/157-1 FUGG) and the
Ministry of Science and Culture (MWK) of the Lower Saxony State.
\end{acknowledgments}

\bibliography{alex_refs,population_refs}

\end{document}